# Surface-state Coulomb repulsion accelerates a metal-insulator transition in topological semimetal nanofilms

**Authors**
S. Ito[1*†], M. Arita[2], J. Haruyama[1], B. Feng[3], W.-C. Chen[4], H. Namatame[2], M. Taniguchi[2], C.-M. Cheng[5], G. Bian[6], S.-J. Tang[4,5], T.-C. Chiang[7], O. Sugino[1], F. Komori[1], I. Matsuda[1*]

**Affiliations**
1. Institute for Solid State Physics (ISSP), The University of Tokyo, Japan
2. Hiroshima Synchrotron Radiation Center (HSRC), Hiroshima University, Japan
3. Institute of Physics, Chinese Academy of Sciences, China
4. Department of Physics and Astronomy, National Tsing Hua University, Republic of China
5. National Synchrotron Radiation Research Center (NSRRC), Republic of China
6. Department of Physics and Astronomy, University of Missouri, USA
7. Department of Physics and Frederick Seitz Materials Research Laboratory, University of Illinois at Urbana-Champaign, USA

*emails of corresponding authors: suguru.ito@physik.uni-marburg.de, imatsuda@issp.u-tokyo.ac.jp
†present address: Department of Physics, Philipps-University of Marburg, Germany

**Abstract**
The emergence of quantization at the nanoscale, the quantum size effect (QSE), allows flexible control of matter and is a rich source of advanced functionalities. A QSE-induced transition into an insulating phase in semimetallic nanofilms was predicted for bismuth a half-century ago and has regained new interest with regard to its surface states exhibiting non-trivial electronic topology. Here, we reveal an unexpected mechanism of the transition by high-resolution angle-resolved photoelectron spectroscopy combined with theoretical calculations. Anomalous evolution and degeneracy of quantized energy levels indicate that increased Coulomb repulsion from the surface states deforms a quantum confinement potential with decreasing thickness. The potential deformation strongly modulates spatial distributions of quantized wave functions, which leads to acceleration of the transition beyond the original QSE picture. This discovery establishes a complete picture of the long-discussed transition and highlights a new class of size effects dominating nanoscale transport in systems with metallic surface states.

**Teaser (127 characters)**
Increased Coulomb repulsion from surface states significantly modulates quantum confinement in topological semimetal nanofilms.

**MAIN TEXT**

**Introduction**

Quantized electronic states generated by the quantum size effect (QSE) in nano-confined systems enable unique tunability for a wide range of phenomena such as superconductivity (*1*), light-matter interaction (*2*), and non-equilibrium carrier dynamics (*3*). Modulations of the band gap and the density of states further improve functionalities in catalysts (*4*) and information devices (*5*). From a technological point of view, quantization inevitably affects any electronic system fabricated at the nanoscale. One of the most well-known examples is a QSE-induced metal-insulator transition, whose essence is illustrated for a film geometry in Figs. 1 (A and B). When semimetallic bulk bands are quantized, the valence-band top and the conduction-band bottom no longer cross the Fermi level ($E_F$), and the system enters an insulating phase. In the case of a system possessing metallic surface states as typically observed in topological materials, the transition is marked by the disappearance of conducting channels in the film interior, and thereafter electric current flows only through the surfaces.

The transition was first predicted a half-century ago on bismuth (Bi) (*6*). A Bi single crystal is a typical semimetal with small carrier pockets and three-dimensional (3D) Dirac dispersions (*7*), which generate unusual magneto-transport responses (*8-10*). Moreover, due to the large spin-orbit coupling, Bi surfaces host spin-polarized metallic states (Figs. 1, A, B, and E) that have been intensively examined in the context of electronic topology (*11-16*). Although theoretical calculations tended to predict trivial band topology (*11, 12*), angle-resolved photoelectron spectroscopy (ARPES) experiments have detected electronic band structures exhibiting nontrivial topology (*13, 14*). A very recent experiment with scanning tunneling microscopy (STM) further supported the presence of a strong topological phase (*15*) and revealed with systematic band structure calculations that pure Bi lies very close to a phase boundary between the strong topological and the higher-order topological phases (*16*). Furthermore, recent STM experiments under strong magnetic fields identified surface Landau levels (*17*) and even a nematic quantum Hall liquid phase formed by the Bi surface states (*18, 19*). The QSE-driven metal-insulator transition in Bi nanofilms, historically called a semimetal-semiconductor transition, originally received great attention as a nanoscale pathway for achieving a substantial thermoelectric figure of merit (*20*) and is now of interest for enhancing the surface-state-induced exotic phenomena.

Evidence of the metal-insulator transition on Bi films was obtained only in this decade by transport measurements on epitaxially grown samples (*21-23*). Initially, measurements were performed *ex situ* on samples with a protective capping layer (*21, 22*), which provoked controversy (*24, 25*) in view of possible contributions from surface oxidization (*26*). The latest experiment finally used *in situ* conditions and concluded that atomically thin Bi films indeed lose conducting channels derived from the quantized bulk states (*23*). Nevertheless, a recent ARPES measurement on Bi films grown exactly in the same condition detected a bulk-derived envelope crossing $E_F$ in atomically thin regions (*27*), in clear contrast to the transport results showing only the interior-insulating phase below a threshold thickness (*21-23*). Although this strange contradiction between metallic and insulating signatures observed in completely the same system implies the presence of an intriguing mechanism, essential quantization information was lacking in previous experiments.

Here, using high-resolution ARPES on high-quality Bi nanofilms, we report the first direct observation of the metal-insulator transition with all the quantized energy levels resolved on the films exhibiting a macroscopically insulating phase. Visualization of anomalous level evolution contrasted with tight-binding simulations highlights an additional mechanism beyond simple QSE. The high-resolution ARPES also detects unexpected degeneracy of top two quantized energy levels, which completely breaks a standard quantization rule. Furthermore, our systematic density-functional-theory (DFT) calculations reveal that the level degeneracy is gradually induced by shifting the whole band structure and accompanies transformation of the bulk-derived wave functions into surface-localized ones. This tendency is totally opposite to a well-known hybridization effect between top and bottom surface states and reconciles the contradiction among the previous experiments in an unprecedented manner. These unusual modulations of quantized bulk states can be fully explained only when we consider deformation of a quantum confinement potential, which is triggered by enhanced effects of Coulomb repulsion with decreasing system size, centering on a size-independent contribution from the surface states. The present study not only solves the serious controversy on the transition discussed for half a century but also introduces the novel size effect which can be universally present in a system with metallic surface states, typically topological materials.

**Results**

**Direct observation of a QSE-induced metal-insulator transition**

Figures 1 (F and G) show Fermi surfaces and band structures measured for a 14 bilayer (BL) Bi film (1 BL = 3.93 Å). Two spin-polarized surface-state bands and quantized bulk bands are distinctly observed in good agreement with those calculated in Fig. 1E. In Figs. 1 (H and I), we systematically follow the evolution of the band structures around the hole and the electron pockets with increasing thickness. The valence-band top at $\bar{\Gamma}$ is resolved together with all the quantized bands and is located well below $E_F$ in atomically thin films. Interestingly, this contrasts with the previous ARPES result (*27*) and is attributed to a finite lattice strain effect, as discussed later. The conduction-band bottom at $\bar{M}$ shifts upwards with decreasing thickness, and the bulk band edges no longer cross $E_F$. To provide more direct evidence, we shifted the whole band structure by *in situ* electron doping and reconstructed band dispersions above $E_F$ (see Supplementary Section A). As far as we know, this is the first direct observation of a metal-insulator transition with all the related electronic states resolved on a thin film system which exhibits macroscopically insulating transport. The situation is in clear contrast to studies on atomically thin films of bulk-conducting topological insulators (*28-30*).

**Anomalous evolution of quantized energy levels observed in the atomically thin regime**

Furthermore, we follow the evolution of the quantized energy levels to investigate mechanisms of the transition, as shown with energy distribution curves (EDCs) at $\bar{M}$ and $\bar{\Gamma}$ in Figs. 2 (A and B). The energy position of each level is described by a boundary condition of confined wave functions (*31*):

$$2k_z D + \Phi = 2\pi(n-1) \qquad (1)$$

where $k_z$ is a wavenumber in the surface-normal direction; $D$ is a film thickness; $\Phi$ is the total phase shift $\phi_{top} + \phi_{bottom}$, as illustrated in Fig. 1B. Eq. (1) tells us that the wavenumber is proportional to the inverse thickness:

$$k_z = \left\{\pi(n-1) - \frac{\Phi}{2}\right\}\frac{1}{D} \qquad (2)$$

Therefore, a plot of quantized energy levels vs. inverse thicknesses depicts a bulk band dispersion perpendicular to the film surface. At $\bar{M}$, the corresponding direction has the Dirac dispersion of Bi (Fig. 1, C and D). Figure 2C shows that evolutions of the peaks are perfectly reproduced by linear functions based on Eq. (2) and a constant phase shift value $\Phi$ obtained in ref. (*14*). However, at $\bar{\Gamma}$ with a parabolic surface-normal dispersion, the peak evolutions are far from the predictions and can be fitted only by a parabolic function with an additional exponential term, as shown in Fig. 2D (see Supplementary Section B for the details of the fitting). This situation becomes more evident when compared to tight-binding calculations with a slab geometry (*32, 33*) in Fig. 2E, where only pure QSE is implemented. In this case, all the peak evolutions in Fig. 2F are consistently described by parabolic functions, which again supports the validity of Eq. (2) with constant $\Phi$. The striking contrast between these two cases indicates that an additional size effect beyond a primitive QSE is driving this anomalous evolution in atomically thin regions. This leads to enhancement of the critical thickness where the top quantized level crosses $E_F$, as highlighted by shaded bars in Fig. 2 (D and F).

**Unexpected level degeneracy and bulk-to-surface transformation**
Another unusual signature appears in the precise measurement of band dispersions. Figure 3A shows second derivative ARPES images for atomically thin films, where both of the two surface-state bands connect to the top quantized bulk band. Because there are states localized at the top and bottom surfaces, these surface-state bands are doubly degenerate in the nearly free-standing film (*34*). This band connection thus indicates that the top quantized level must be quadruply degenerate. That is, $n = 1$ and $2$ quantized levels are almost degenerate (Fig. 3B), which has never been reported in ARPES experiments of thin films.

Nevertheless, the observations can be reproduced by first-principles calculations which systematically shift the entire band structures, as shown in Fig. 3C. The shifts are manually induced by changing lattice parameters, where the in-plane lattice constant is modulated by -2, -1, 0, +1 % from left to right, respectively, with the total volume conserved. Along with the shift toward a higher binding energy, the energy separation between $n = 2$ and $3$ levels increases. In contrast, the separation between $n = 1$ and $2$ levels is reduced; their eventual degeneracy is consistent with Fig.3 (A and B). It is as if the quantized bands split off from the bulk projection and behave independently as surface-state bands. This is supported by examining wave functions for each quantized state in Fig. 3D. In the left panel, the wave functions have envelope shapes expected for $n = 1, 2, 3$ states confined in a quantum well. However, as the unusual level degeneracy appears, wave functions of $n = 1$ and $2$ states are transformed into surface-localized ones. It is well known that electronic states localized on top and bottom surfaces can hybridize and exhibit a bulk-like character in atomically thin films (*29*). In the case of Bi, this hybridization effect was demonstrated only around $\overline{M}$ and do not exist near $\overline{\Gamma}$ because of a short decay length of the surface states therein (*14, 35*). Here, the strikingly opposite behavior to the hybridization effect, the transformation of bulk-derived states into surface-localized ones, is observed for the quantized bulk states around $\overline{\Gamma}$.

**The central mechanism of the present observations**
An important question is what is the central mechanism responsible for the anomalous behaviors. The level degeneracy rules out major contributions of standard size effects related to charge transfer, lattice strain, and recently proposed surface-size effects (*27*), all of which only uniformly shifts or expands the quantized bands and never generates degeneracy. Note that the lattice strain effect corresponds to a uniform compression/expansion of in-plane/out-of-plane lattice constants due to a lattice mismatch with substrates. We can still think of modulation in the interlayer spacing, which can affect Bi band structures to a relatively large extent (*36*). However, this effect is also excluded as the central mechanism of the present observations because our DFT calculations already exhibit the level degeneracy and the bulk-to-surface transformation using a homogeneous Bi slab without any modulation of the interlayer spacing implemented. (See also Supplementary Section C for detailed considerations about the conventional effects.)

Although the transformation of $n = 1$ and $2$ confined wave functions into surface-like states can be viewed as a result of hybridization between the surface and bulk states around $\bar{\Gamma}$, the reason for the degeneracy between these two quantized bulk states is not yet accounted for. More importantly, we need to address the reason why the level degeneracy and the bulk-to-surface transformation are gradually induced by shifting the whole band structure in Fig. 3 (C and D). Because the DFT calculation is performed on a free-standing Bi slab, any substrate or interface effect is strictly excluded. Band shifts induced by the lattice-parameter modulation cannot explain them, either, as confirmed by tight-binding calculations (see Supplementary Section C). An essential difference between the present tight-binding and first-principles approaches is an implementation of Coulomb interaction. In the general framework of DFT, an effective one-body potential $V_{eff}$ and a total charge density $n_{total}$ are determined by a self-consistent cycle reflecting Coulomb interaction via the Hartree functional and the exchange-correlation functional (*37*). Here, the occupation of the surface-state bands increases from left to right in Fig. 3C, which increases their relative contributions to the total charge density $n_{total} = n_{surface} + n_{interior}$: this can modify the effective potential $V_{eff}$ and electronic structures finally obtained.

We further conceive that Coulomb repulsion among electrons tends to compress the total charge distribution toward the film center when the surface contributions increase. The behavior is indeed observed in the present calculations (see Supplementary Section D). In the one-body picture of the DFT framework, the compressed total charge distribution in turn makes an electron feel a potential barrier around the film center and deforms $V_{eff}$ into a double-well-like shape, as illustrated in Fig. 4A. Eigenstates of a double-well potential have nearly degenerate ground states with wave functions localized in each well; this then provides a qualitative explanation of the experimental observations. Whereas the modulation of the electronic states is triggered by an increase of surface-state occupations in the calculation, the positions of the surface-state bands are almost independent of thicknesses in real Bi films (Fig.1, H and I). Nevertheless, reducing the system size increases relative contributions of the surface states to the total charge density, which has essentially the same effect. We also examine this picture by performing a numerical simulation using a 1D Schrödinger equation with an ideal quantum-well potential that is gradually deformed to a double well. Figure 4B shows that this simple model quantitatively reproduces both the anomalous evolution and the level degeneracy when a double-well modulation and the resulting change in a band dispersion (an effective mass) have an exponential thickness dependence (see Supplementary Section E). The insets also demonstrate that the wave functions are indeed transformed to surface-localized ones with decreasing thickness.

Thus, the two essential signatures here, the degeneracy between the top two quantized bulk states and their transformation into surface-localized states, rule out scenarios given by any standard mechanism but are naturally explained by considering the deformation of a quantum confinement potential via Coulomb repulsion from the surface states. One of the most interesting points is that a usually weak effect of Coulomb repulsion and hybridization between the surface and quantized bulk states are enhanced with decreasing system size centering on the presence of the surface states.

**Discussion**

We revisit the strange contradiction among recent studies on the metal-insulator transition in Bi films. The central problem is a metallic envelope of quantized bulk states captured by the previous ARPES of atomically thin Bi films grown on a Si(111) substrate (*27*), whose film interiors were insulating in the transport experiments (*22, 23*). Using the same substrate, we reproduced the ARPES data. As experimentally studied in ref. (*38*), the smaller Si lattice constant relative to that of Ge exerts stronger compressive strain. The stronger strain in Bi/Si films shifts the whole band structure upwards compared to the Bi/Ge case, so that the top quantized level crosses $E_F$ (see Figs. S2C). Nevertheless, as shown in Fig. 3D and in the insets of Fig. 4B, the wave function of this top quantized level behaves like a surface-conducting channel, and the resulting shift of an effective band edge makes the system interior-insulating even in the Bi/Si case, which is consistent with the transport experiments (*21-23*). (See also Supplementary Section F for careful considerations of surface and bulk transport channels in atomically thin films.) Therefore, in addition to the exponential level evolution pushed by the emergent double-well potential, the bulk-to-surface transformation further accelerates the metal-insulator transition. This is a complete picture of the long-discussed problem in Bi, whose mechanism is unexpectedly extended centering on the surface states.

The present conclusion can be generalized to a new class of size effects in any dimension, where increased contributions of one- and two-dimensional edge states with decreasing system size modify an effective one-body potential via Coulomb repulsion. Here, the first experimental signature on a specific system still calls for more detailed theoretical studies in a wide range of materials. Nevertheless, the comprehensive discussions presented above suggest that the effect is likely to be present in a universal system with metallic edge states. A target of the greatest interest will be nanofilms of recently discovered Dirac (*39*), Weyl (*40, 41*), and topological nodal-line semimetals (*42*) that inherently possess metallic edge states and a small number of bulk carriers near $E_F$ inside their point-/loop-like semimetallic nodes. One-dimensional edge/hinge states in two-dimensional (*34, 43*) or higher-order topological insulator phases (*16*) will also be an interesting playground in view of the stronger confinement. Nano-fabrication of such topological materials is an indispensable step for their device applications, in which various size effects including the one discovered here will play an essential role and provide new control parameters for advanced functionalities.

## Materials and Methods
### Sample preparation
Bi(111) films were grown epitaxially on a medium-doped *p*-type Ge(111) wafer cleaned by cycles of Ar$^+$ sputtering and annealing at 900 K. Bi evaporation was performed at room temperature, followed by annealing at 400 K (*38*). To improve the film quality, the Ge substrate was prepared with large domains by fully outgassing the preparation systems and uniformly sputtering and annealing the substrate. The temperature during the Bi deposition was also carefully controlled. The relative accuracy of film thicknesses was precisely controlled using a quartz thickness monitor, and absolute film thicknesses were calibrated by comparing quantized energy levels with previous reports (*34, 44*). The high film quality was also confirmed by comparing sharpness in the photoemission spectra of fine quantization structures (Fig. 1 H and I) with the previous reports. The thinnest thickness (7 BL) was set close to the critical thickness of the Bi(111) structure on a Ge(111) substrate (*38*).

### ARPES measurements
High-resolution ARPES measurements were performed at BL-9A of the Hiroshima Synchrotron Radiation Center (HSRC) and BL-21B1 of the National Synchrotron Radiation Research Center (NSRRC). At BL-9A, a high-intensity unpolarized Xe plasma discharge lamp (8.437 eV) was used for magnifying small energy ranges, in addition to the synchrotron radiation (21 eV) for wide-range observations. The measurement temperature was 10 K, and the total energy resolution was 7 meV for 8.437 eV photons and 12 meV for 21 eV photons. The pressure during the measurements was 10$^{-9}$ Pa. The measurement direction was precisely adjusted using an automated 6-axis rotational controller, and was based on Fermi surfaces that were mapped before every band-structure scan. To directly determine unoccupied band structures, we also performed alkali-metal adsorption following an approach in ref. (*45*) (see Supplementary Section A).

**Tight-binding calculations**
Tight-binding calculations for bulk electronic structures of Bi were based on a framework introduced in ref. (*46*), which considers a $16\times16$ matrix composed of hopping parameters of the $sp^3$ orbitals between first-, second-, and third-nearest neighbor atoms with spin-orbit coupling implemented. Electronic structures in a slab geometry were calculated by extending the bulk matrix to a larger $16N\times16N$ matrix, where $N$ is the number of bilayers, as demonstrated in ref. (*32, 33*). The lattice parameters describing a Bi rhombohedral unit cell and the hopping parameters were taken from ref. (*46*). The surface potential term was also introduced using formulations and parameters in ref. (*32*).

**First-principles calculations**
Density functional theory (DFT) calculations (*37*) were performed with the ABINIT code (*47*). The gradient approximation was selected for the exchange-correlation functional, and a norm-conserving pseudopotential was used in which spin-orbit coupling was implemented. A free-standing Bi slab with a 14 BL thickness was used, and the length of a vacuum region was set to 8 BL (31.5 Å). Lattice parameters of the slab were fixed to the experimental values (*46*), where an in-plane lattice constant is 4.55 Å and intra-bilayer and inter-bilayer spacings are 1.59 Å and 2.34 Å, respectively. A grid for $k$-point sampling as $9\times9\times1$ was used. Convergences with respect to an energy-cutoff, the $k$-point sampling, and the vacuum length were confirmed for obtained band structures. We also cross-checked the results using the QUANTUM-ESPRESSO code (*48*) with similar conditions.

**Model calculations**
A numerical simulation was performed using a 1D Schrödinger equation with an electron mass extracted from the tight-binding band structure. Eigenvalues were calculated by taking the Fourier transform of a wave function $\varphi(z)=\sum_{i=1}^{\infty}\sqrt{\frac{2}{L}}\alpha_i \sin(\frac{i\pi}{L}z)$ where $L$ is a maximum length scale of the simulation (*49*). A finite-height single-well potential was set inside this length scale. A double-well deformation of the single well was introduced with an exponential dependence on thickness. To manually simulate modulation of a band dispersion accompanying the potential deformation, an exponential dependence was also installed for the electron effective mass. All the parameters were adjusted to best fit the experimental data. Further details are described in Supplementary Section E.

**Acknowledgments**

**General**: We thank Prof. T. Hirahara and Prof. M. Horn-von Hoegen for valuable discussions on the growth of Bi films. We also thank Prof. T. Kato, Dr. Y. Konishi, and Dr. T. Miyamachi for fruitful discussions on edge states in a confined geometry. We acknowledge Prof. T. Aono, Dr. Y. Ohtsubo and Dr. R. Yukawa for advice on theoretical calculations. We thank Dr. Alan Burns from the Edanz Group for checking a draft of the manuscript.

**Funding:** The ARPES measurements were performed with the approval of the Proposal Assessing Committee of HSRC (Project No. 15-A-38, 17AU025, and 18AU008) and the Proposal Assessing Committee of NSRRC (Project No. 2015-2-090-1). I.M. acknowledges support by JSPS under KAKENHI Grant No. 18H03874. T.C.C. acknowledges support from the U.S. National Science Foundation under grant number DMR-17-09945. S.I. acknowledges support by JSPS under KAKENHI Grant No. 17J03534. S.I. was also supported by JSPS through the Program for Leading Graduate Schools (ALPS).

**Author contributions:** S.I. and I.M. planned the experimental project. S.I. fabricated thin film samples, conducted ARPES experiments, and analyzed the data. M.A., B.F., W.C.C., C.M.C. supported the ARPES experiments. S.I. performed tight-binding and model calculations. S.I. also performed first-principles calculations with support by G.B., J.H., and O.S. S.I., O.S., F.K., and I.M. organized discussions on the results, and S.I. wrote the paper. All authors contributed to the discussions and commented on the manuscript.

**Competing interests:** The authors declare that they have no competing interests.

**Data and materials availability:** All data needed to evaluate the conclusions in the paper are present in the paper and/or the Supplementary Materials. Additional data are available from the authors upon request.

**The Supplementary Material PDF file includes:**

- section A. Direct observation of the conduction-band bottom located above $E_F$ at $\bar{M}$
- section B. Details of fitting functions in Fig. 2D of the main text
- section C. Additional considerations of standard size effects
- section D. Changes of a total charge distribution in the first-principles calculations
- section E. Details of model calculations in Fig. 4B of the main text
- section F. The distinction between surface and bulk states in atomically thin Bi films
- fig. S1. Direct observation of the conduction-band bottom located above $E_F$ at $\bar{M}$.
- fig. S2 No substrate doping dependence of quantized energy levels at $\bar{\Gamma}$
- fig. S3. A finite effect caused by lattice strain from substrates.
- fig. S4. Experimental evaluation of minor contributions from all standard size effects.
- fig. S5. Changes of a total charge distribution in the first-principles calculations.
- fig. S6. Details of model calculations in Fig. 4B of the main text.
- References 6, 14, 23, 27, 29, 34, 35, 38, 45, 46, and 49

**Figures**

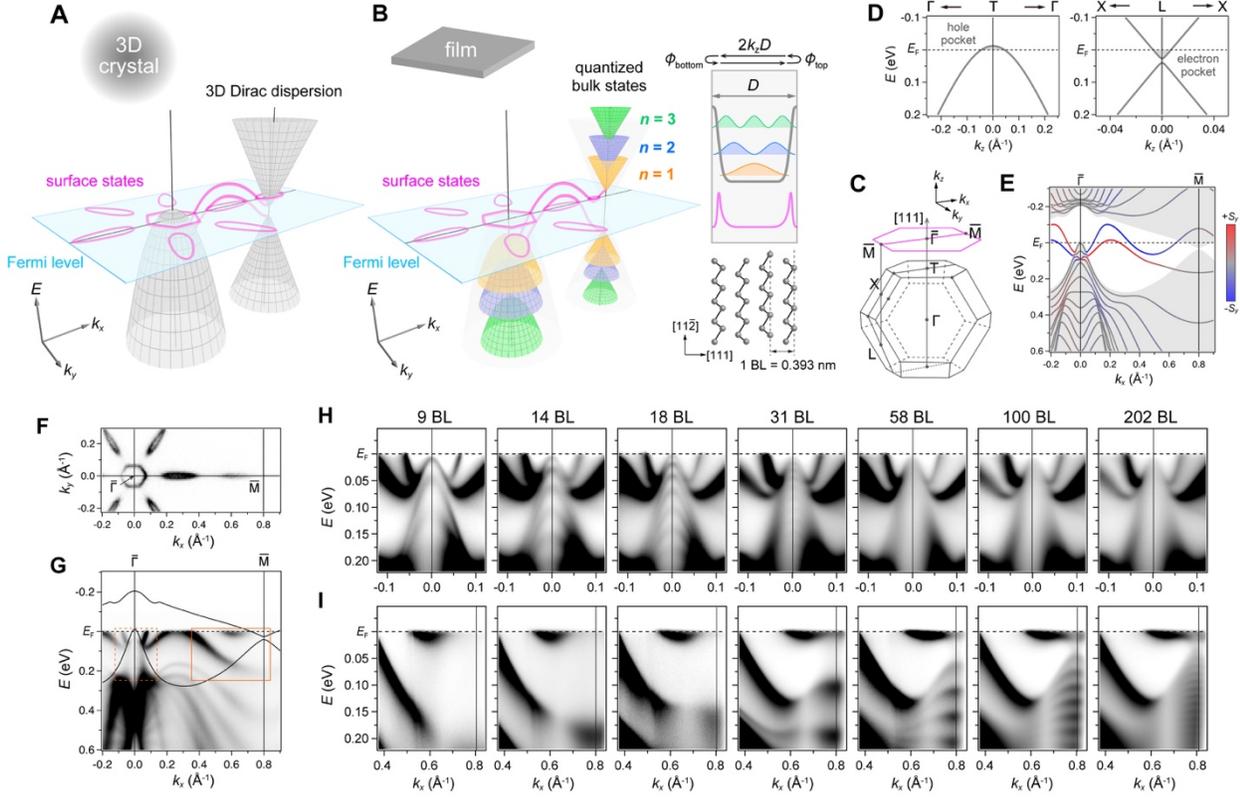

**Fig. 1. Direct observation of a QSE-induced metal-insulator transition.** (**A** and **B**) Schematic of the metal-insulator transition in a Bi nanofilm. While 3D bulk states become quantized and gapped across the Fermi level, metallic surface states remain intact. The inset of (**B**) depicts wave function characters of the surface states and the quantized bulk states formed inside a film with thickness $D$. Circulating arrows show a boundary condition for the latter, where $\phi_{top} + \phi_{bottom}$ are phase shifts when reflected at top and bottom surfaces. The drawing below also illustrates a bilayer (BL) structure of Bi. (**C**) Bulk and surface Brillouin zone of Bi in the [111] direction. (**D**) Bi bulk band structures calculated around the hole and electron pockets using a tight-binding method. (**E**) Band structures obtained by a tight-binding calculation for a 14 BL Bi(111) slab. The color scale shows in-plane spin-polarization of each state at the top surface in the direction perpendicular to $\overline{\Gamma M}$ (**F** and **G**) Experimental Fermi surfaces and band structures measured along $\overline{\Gamma M}$ on a 14 BL Bi(111) film grown on a Ge(111) substrate. Shaded areas in (**E**) and solid curves in (**G**) show calculated bulk projections. (**H** and **I**) Experimental band structures magnified inside dashed and solid boxes in (**G**), respectively, for each thickness.

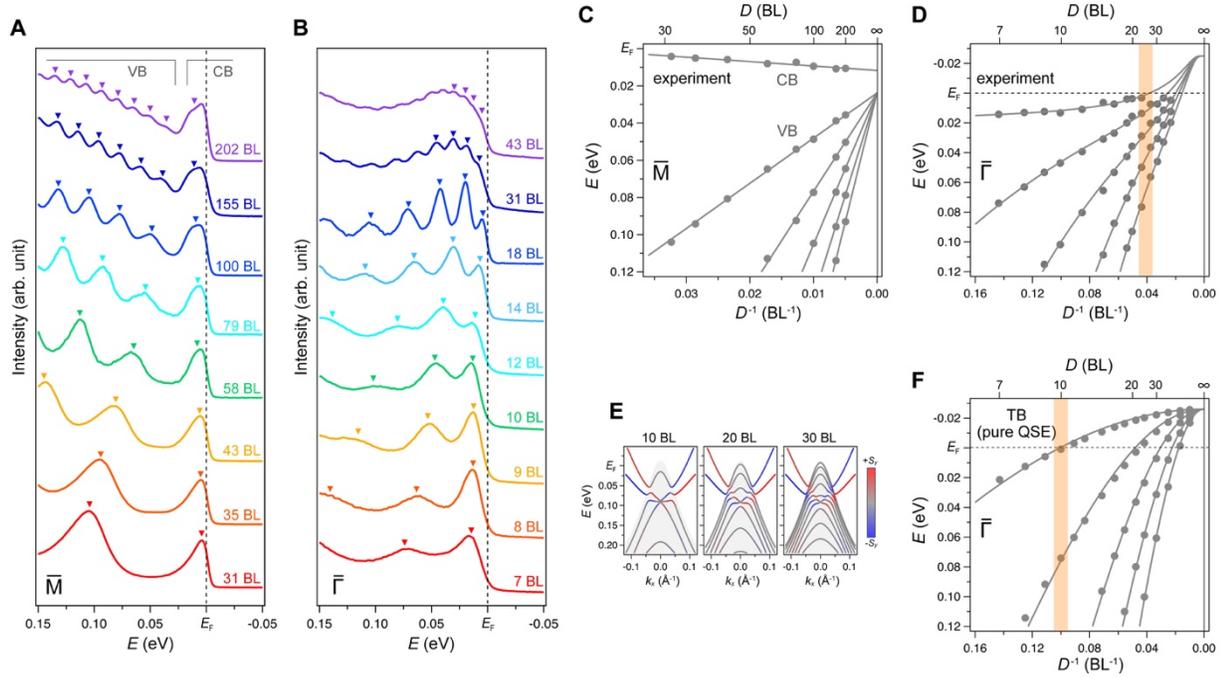

**Fig. 2. Anomalous evolution of quantized energy levels observed in the atomically thin regime.** (**A** and **B**) Evolution of energy distribution curves (EDCs) extracted at $\overline{\mathrm{M}}$ and $\overline{\Gamma}$ for each thickness. Markers show peak positions determined by Lorentzian fitting. Peaks belonging to the conduction band (CB) and the valence band (VB) are also denoted in (**A**). (**C**) Peak positions at $\overline{\mathrm{M}}$ vs. inverse thicknesses of each film. Linear functions are based on Eq. (2) with a surface-normal band dispersion and a phase shift obtained experimentally (*14*). (**D**) The same as (**C**) for peak positions at $\overline{\Gamma}$. Solid curves are fitting functions discussed in the main text and Supplementary Section B. A shaded bar highlights a thickness where the top quantized level crosses $E_\mathrm{F}$. (**E**) Evolution of band structures around $\overline{\Gamma}$ calculated by a tight-binding method for Bi slabs. The color scale is the same as in Fig. 1D. (**F**) The same as (**D**) for energy levels obtained by the tight-binding calculations that reflect only pure QSE.

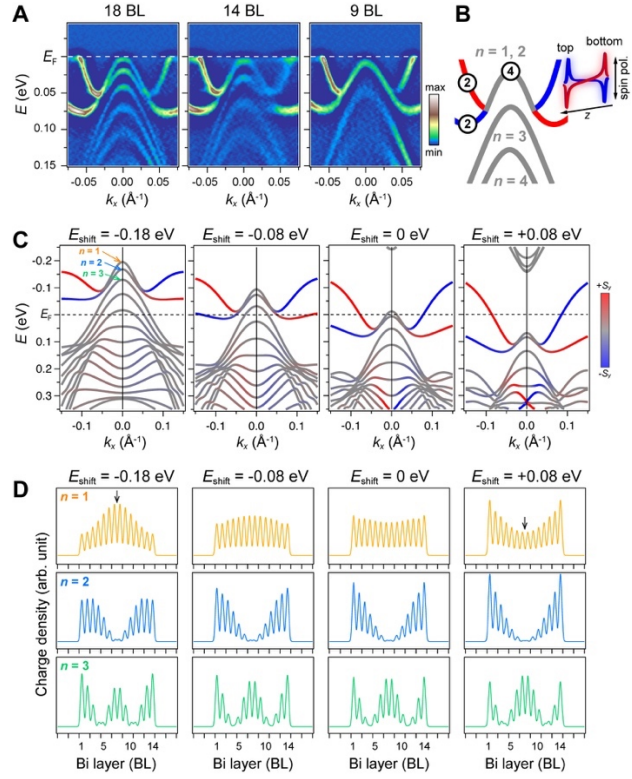

**Fig. 3. Unexpected level degeneracy and bulk-to-surface transformation.** (**A**) Second derivative ARPES images for highlighting peak positions around $\bar{\Gamma}$ in atomically thin Bi films. (**B**) Schematic of unusual band connections between the top quantized band and the surface-state bands. The number of states in each band is explicitly shown with illustrations of wave functions localized near the top and bottom surfaces. (**C**) Evolution of band structures shifted in energy by tuning lattice parameters in first-principles calculations of a 14 BL Bi slab. The color scale depicts spin-polarization in the same manner as in Fig. 1D. (**D**) Squared values of wave functions for $n = 1, 2, 3$ states at $\bar{\Gamma}$ in (**C**). The horizontal axis corresponds to the direction perpendicular to the slab surface, where in-plane contributions are integrated.

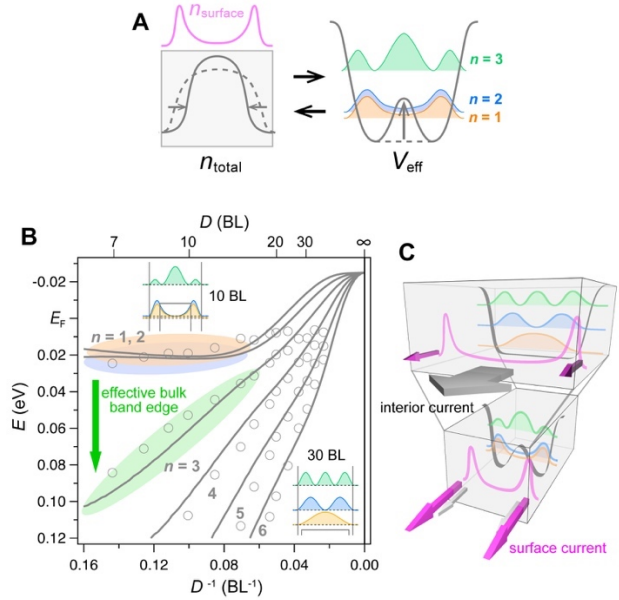

**Fig. 4. Surface-state Coulomb repulsion induces deformation of a quantum confinement potential.** (**A**) Schematic of a self-consistent cycle for a total charge density $n_{total}$ and an effective confinement potential $V_{eff}$, which is triggered by increased contributions of surface states and concomitant compression of the total charge distribution via Coulomb repulsion. Energy levels and wave function characters of eigenstates in a double-well potential are also illustrated. (**B**) Solid curves: evolution of quantized energy levels simulated using an ideal quantum-well potential subject to a gradual double-well modulation. Markers show experimental data corrected for standard size effects (see Supplementary Section D). The insets show squared values of wave functions that are calculated for 10 BL and 30 BL cases. (**C**) Schematic representation of the novel size effect dominating nanoscale electronic transport in a system with metallic surface states, centering on a size-independent contribution from the surface states.
14

# Supplementary Materials

## A. Direct observation of the conduction-band bottom located above $E_F$ at $\bar{\mathrm{M}}$

In the main text, it is confirmed that the conduction-band bottom at $\bar{\mathrm{M}}$ is located above $E_F$ in atomically thin Bi films based on the disappearance of the ARPES intensity. Here, we demonstrate the more direct evidence. To observe the unoccupied band structures by ARPES, we doped electrons into the bands by alkali-metal (Cs) adsorption. Cs adsorption was performed using a commercial alkali-metal dispenser. Figure S1 (A and B) show wide-range ARPES images for pristine and Cs-adsorbed films, respectively, with extracted peak positions of two surface-state bands highlighted by markers. As discussed in ref. (*45*), Cs atoms cause state-selective electron doping connected to the extent of surface localization of each state of a substrate. We first evaluate the quantitative relation between them for reconstructing the original band structures. Figure S1C shows band structures with surface charge densities mapped on each state, which is calculated following procedures in ref. (*45*). In Figs. S1 (D and E), we compare wavenmber dependence of band shifts between the pristine and Cs-adsorbed cases (markers) with that of the surface charge densities (solid curves). The surface charge densities are rescaled manually to quantitatively fit the band shifts in both SS2 and SS1 cases. Because SS1 peak positions in a pristine case are located near $E_F$ and can be strongly affected by the Fermi-Dirac function, we only consider a peak corresponding to the band bottom as for SS1, which is highlighted by a pink marker. The obtained scaling relation enables us to evaluate band shifts even for $(E,k)$ points without experimental data in the pristine case. Using the evaluated band shifts and peak positions in the Cs-adsorbed case, we can reconstruct an SS1 band structure hybridizing with the conduction-band bottom in the unoccupied region, as shown in Fig. S1F.

## B. Details of fitting functions in Fig. 2D of the main text

In the main text, we fitted the evolution of quantized energy levels extracted at $\bar{\Gamma}$ using a following function:

$$E = ak_z^2 + b + c\exp(-D/d) \quad (S1)$$

Using Eq. (2) of the main text and an approximation of a constant phase shift $\Phi$, we can rewrite the formula as

$$E_n = A_n(1/D)^2 + b + c_n\exp(-D/d) \quad (S2)$$

where $n$ is an index of a quantization number. (Note that at this point we are not aware of the new size effect causing degeneracy of $n = 1$ and 2 demonstrated in Fig. 3 of the main text and that we assign the quantization numbers regularly from top to bottom.) In the actual fitting process, the value of $b$, the valence-band top of the hole pocket, is fixed to -0.025 eV, a value determined from a semimetallic band overlap in ref. (*46*) and the conduction-band bottom experimentally obtained in ref. (*14*). As one can guess from a lack of the index $n$, here $d$ is also assumed to be common to all the quantized levels, which significantly improves consistency in the whole fitting. In short, the solid curves in Fig. 2D of the main text are obtained by a fitting process using Eq. (S2) with $A_n, c_n (n=1,2,3,4,5)$ and $d$ treated as independent parameters.

## C. Additional considerations of standard size effects

In the main text, we simply excluded major contributions of any standard size-dependent effect based on the emergent degeneracy of quantized energy levels in our DFT calculations. In this section, we introduce additional discussions of conventional effects: (i) no substrate doping dependence of quantized energy levels at $\bar{\Gamma}$, (ii) a finite effect caused by lattice strain from substrates, and (iii) experimental evaluation of minor contributions from all standard size effects.

### *C.1. No substrate doping dependence of quantized energy levels* at $\bar{\Gamma}$

Figures S2 (A, B, and C) show band structures measured around $\bar{\Gamma}$ on Bi films grown on Ge(111) substrates with three different doping conditions: *p*-type medium doped, *n*-type medium doped, and *p*-type highly doped. If there exists a charge transfer channel from a substrate to a Bi film, the difference in the doping conditions will have direct influence on the number of migrating carriers and their signs. However, the results show completely identical band structures. By superimposing their EDCs extracted at $\bar{\Gamma}$ in Fig. S2D, we see that their quantized energy levels also show excellent quantitative agreements. This result tells us that even a phase shift at an interface is not affected by the substrate doping conditions and clearly excludes the presence of charge transfer in this system, which is consistent with a free-standing character of Bi films discussed previously (*34*).

### *C.2. A finite effect caused by lattice strain from substrates*

Figure S3A shows the dependence of the VBM position at $\bar{\Gamma}$ on in-plane lattice strain. We note that compressive in-plane strain rather drives Bi away from the insulating phase. We can confirm a consistent tendency both for surface bands and quantized bulk bands of Bi films in Fig. S3B. Figure S3C shows experimental band structures measured on 10 BL Bi films respectively grown on Ge(111) and Si(111) substrates, where we can see a finite energy difference between these two cases. This shift makes the top quantized band cross $E_F$ for the Bi/Si case (Fig. S3D) and nicely reproduces the previous ARPES result (*27*). Because the in-plane lattice constant of Si is ~5 % smaller than that of Ge, the energy shift is consistent with the compressive-strain effect. (The stronger compressive strain was also experimentally demonstrated in ref. (*38*).) Although it turns out that the strain effect has a finite contribution, this effect cannot explain the unusual degeneracy of quantized energy levels as we can clearly see in Fig. S3B. In addition, we can also exclude a major contribution of this effect to the metal-insulator transition in Bi/Ge films, based on its totally opposite behavior. Because of the smaller lattice constant of Ge than that of Bi, an in-plane lattice constant of a Bi film on a Ge substrate receives stronger compressive strain with decreasing thickness (*38*). This tendency counteracts the metal-insulator transition and is contrary to the acceleration of the transition observed in the present study (see shaded bars in Figs. 2 D and F of the main text).

### *C.3. Experimental evaluation of minor contributions from all standard size effects*

In addition to the charge transfer and the lattice strain effect, hybridization of top and bottom surface states (*29, 34, 35*) and a charge neutrality condition between surface and bulk carriers (*27*) are also suggested as size-dependent effects in an atomically thin film. As for Bi films, the former affects only states near $\bar{M}$ because of the longer decay length of the surface-state wave functions (*14*). The latter simply emerges as a shift of the Fermi level. Therefore, all the size effects related to quantized levels at $\bar{\Gamma}$ just uniformly shift or compress band structures, as mentioned in the main text.

Because binding energies of surface states depend on thickness only via the standard size effects, we can experimentally evaluate minor contributions of these effects just by checking the evolution of peak positions marked in Fig. S4A. We plot the extracted peak positions as relative energy shifts in Fig. S4B. The result clearly shows a relaxation with increasing thickness, which is nicely fitted by a double exponential function. This energy shift corresponds to a sum of contributions from all the standard size effects. By using this energy shift, we can subtract contributions of these standard size effects from the experimental evolution of quantized energy levels. If everything can be explained by a combination of QSE and the standard size effects, the corrected evolution of quantized levels must follow a simple parabolic function as in the ideal case of Fig. 2F of the main text. However, as shown in Fig. S4C, it is obvious that the energy shift is still too small (~10 meV) to correct the experimental evolution to a parabolic one. This analysis supports our conclusion in the main text from a different and purely experimental viewpoint.

**D. Changes of a total charge distribution in the first-principles calculations**
In the main text, we discuss deformation of an effective one-body potential driven by compression of a total charge distribution toward a film center. This tendency is indeed observed in the first-principles calculations performed in the present study (Fig. 3 C and D of the main text). Figure S5 (A and B) show the corresponding changes in the total charge distribution along the surface-normal direction where the in-plane contributions are integrated. We notice that charge densities inside slab interiors gradually increase together with the appearance of the level degeneracy and the bulk-to-surface transformation of wave function characters. We must note that these behaviors are originally induced by reducing the in-plane lattice constant. Nevertheless, an important fact is that charge densities near slab edges decrease, which cannot be explained by mere compression of the slab. This feature clearly shows that the changes in the total charge density can only be understood based on a self-consistent DFT cycle, where many-body electronic correlations play an essential role.

One may think that the more straightforward evidence of the picture proposed in the present study is a consistent change in the effective one-body potential itself. We can quickly check the change from the calculation results, but we must be careful of the fact that the effective confinement potential determining the envelope shapes of the wave functions in Fig. 3D of the main text is not necessarily the same as the one-body potential determined in the DFT calculations (Kohn-Sham potential). The difference becomes evident when we note that the quantized energy levels at $\bar{\Gamma}$ are formed inside the hall pocket, whose description requires an inverted confinement potential. We judge that full construction of such a local pseudopotential is beyond the scope of the present study as the first detection of the experimental signature. We instead apply a simplified simulation using a 1D Schrödinger equation, as described in the following section.

**E. Details of model calculations in Fig. 4B of the main text**
In this section, we introduce details of model calculations performed in Fig. 4B of the main text to simulate the effects of the double-well modulation of a confinement potential. The calculation is conducted based on a 1D Schrödinger equation:

$$-\frac{\hbar^2}{2m_*}\frac{d^2}{dz^2}\psi(z) + V(z)\psi(z) = E\psi(z) \tag{S3}$$

where $m_*$ is an effective mass. Following the approach of ref. (*49*), we take the Fourier transform of a wave function

$$\psi(z) = \sum_{i=1}^{M} \sqrt{\frac{2}{L}} \alpha_i \sin\left(\frac{i\pi}{L} z\right). \tag{S4}$$

$L$ is a maximum length scale of the simulation and $M$ a maximum number of Fourier bases. Then we can transform Eq. (S3) to

$$\sum_j H_{ij} \alpha_j = E_i \alpha_i. \tag{S5}$$

$H$ is a $M \times M$ matrix, whose element is given by

$$H_{ij} = \frac{\hbar^2}{2m_*}\left(\frac{i\pi}{L}\right)^2 \delta_{ij} + \frac{2}{L}\int_0^L dz V(z) \sin\left(\frac{i\pi}{L}z\right)\sin\left(\frac{j\pi}{L}z\right) \tag{S6}$$

where $\delta_{ij}$ is the Kronecker delta. Eigenvalues and eigenfunctions are obtained by solving the secular equation.

Wave functions described by Eq. (S4) necessarily satisfy boundary conditions of $\psi(z=0,L)=0$, corresponding to confinement by infinitely high potential walls. To express imperfect confinement realized in real materials, we set an ideal well potential with a finite height $V_{SW}$ inside this infinitely high well. We also introduce a double-well modulation characterized by a height $V_{DW}$. The situation is illustrated in Fig. S6A. A size-dependent potential modulation and the resulting deformation of a band dispersion (an effective mass) are implemented by manually introducing an exponential dependence on a thickness $D$ into $V_{DW}$ and $m_*$:

$$V_{DW}(D) = V_{DW0} \exp(-D/d_V) \tag{S7}$$
$$m_*(D) = m_{*0} + \Delta m_* \exp(-D/d_m) \tag{S8}$$

where $m_*$ is fixed to $0.7 m_e$, a value extracted by fitting a Bi band structure around T point (Fig. 1D of the main text). The values of $V_{SW}$, $D_s$, $V_{DW0}$, $d_V$, $\Delta m_*$, and $d_m$ are adjusted to best fit the corrected experimental evolution (Fig. S4C), under a constraint that $n=1$ and 2 energy levels are almost degenerated. The determined values are $V_{SW}=0.40$ eV, $D_s=2.0$ BL, $V_{DW0}=0.18$ eV, $d_V=10$ BL, $\Delta m_*=5.3 m_e$, and $d_m=10$ BL.

The evolution of the calculated energy levels is shown in Fig. S6B together with the experimental data. To calculate a case very close to the bulk limit, the total length scale $L$ is set to 500 BL, which requires $M=800$ bases for converged calculations in atomically thin regions. Squared values of wave functions for 30 BL, 10 BL, and 7 BL cases are depicted in Figs. S6 (C, D, and E), respectively. Although decay parameters are set to above values, the resulting thickness range where the modulation of quantized energy levels and wave function characters becomes evident is roughly estimated as 20-30 BL from Figs. S6 (B, C, D, and E), which is shifted by 10-20 BL compared to a critical thickness in Fig. 2F. Because the tight-binding simulation reflects only pure QSE, this directly corresponds to acceleration of the transition in relation to the original prediction half a century ago (*6*).

**F. The distinction between surface and bulk states in atomically thin Bi films**

In the present study, we interpreted all the results considering surface states and quantized bulk states formed in Bi thin films. However, careful readers may be concerned that the distinction between surface and bulk states can be ambiguous in atomically thin films owing to the hybridization between top and bottom surface states. The transport experiment in ref. (*23*) indeed detected a signature of surface states possessing finite contributions in the film interior. Nevertheless, the experiment succeeded in separating thickness-dependent/independent transport channels and demonstrated the disappearance of the former channel in atomically thin Bi films. In contrast, the ARPES measurements detected the top quantized level crossing $E_F$ in a 10 BL Bi film grown and measured in exactly the same conditions (usage of a Si(111) substrate, growth methods, and ultrahigh vacuum), as shown in Fig. S3C. Because the disappearing channel is clearly derived from thickness-dependent quantized bulk states, the transport and the ARPES results seriously contradict each other as long as we assume a purely quantum-well character of the electronic state crossing $E_F$. (See the leftmost panel of Fig. 3D.) By considering that the state crossing $E_F$ loses a character of quantized bulk states and behaves as effective surface states, these two experimental results can be consistently understood. In other words, the fact that the *in situ* transport experiment detected the disappearance of the thickness-dependent channel in this thickness range is firm evidence that the surface-localized state in the rightmost panel of Fig. 3D exhibits a macroscopic response that is distinguishable from that of typical quantized bulk states. In general, the hybridization effect is inevitable and whether surface states are well defined or not must always be tested in individual film systems approaching the two-dimensional limit. A key criterion is the detectability of a unique macroscopic response, as demonstrated here.

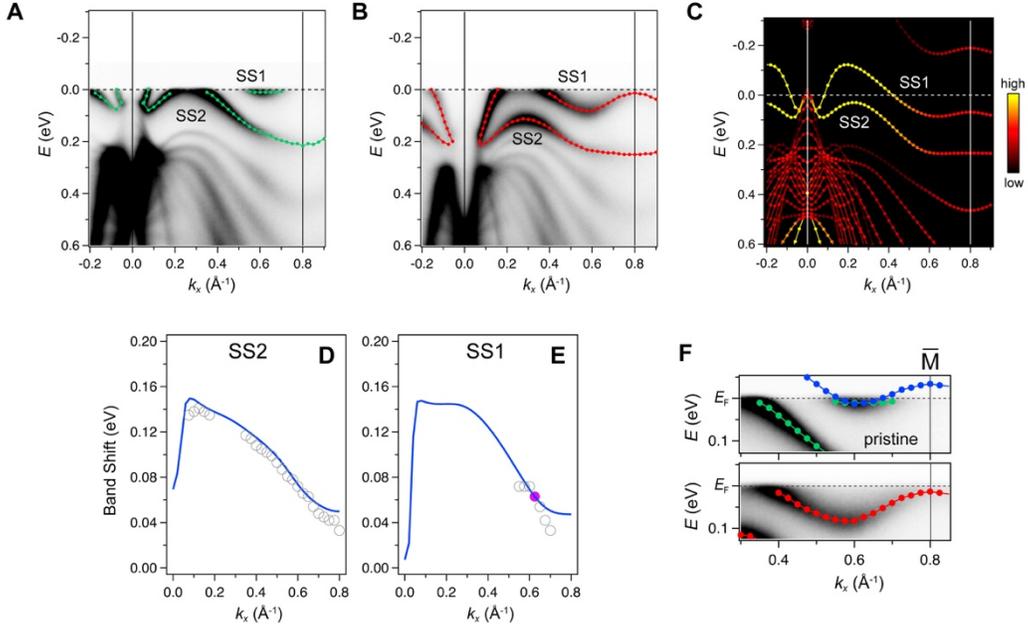

**Fig. S1. Direct observation of the conduction-band bottom located above $E_F$ at $\overline{\mathrm{M}}$.** (**A** and **B**) Wide-range ARPES images measured on pristine and Cs-adsorbed films, respectively, with a thickness of 14 BL. Markers show peak positions extracted for two surface-state bands, SS1 and SS2. (**C**) Band structures with surface charge densities mapped on each $(E,k)$ point, which is calculated following procedures in ref. (*45*). (**D** and **E**) Wavenumber-dependent SS2 and SS1 band shifts between the pristine and Cs-adsorbed cases directly superimposed by wavenumber-dependent surface charge densities, which are manually rescaled to best fit the band shifts. (**F**) Comparison of original ARPES results and an SS1 band structure reconstructed using the result in (**E**) and peak positions of the Cs-adsorbed film.

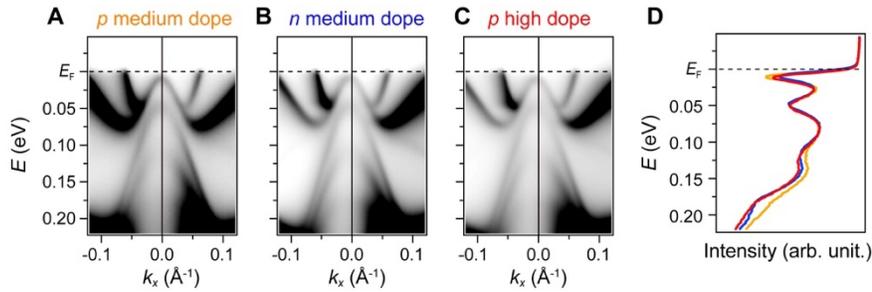

**Fig. S2. No substrate doping dependence of quantized energy levels at $\overline{\Gamma}$.** (**A**, **B**, and **C**) Band structures around $\overline{\Gamma}$ measured on Bi films grown on Ge(111) substrates with three different doping conditions: *p*-type medium doped, *n*-type medium doped, and *p*-type highly doped. (**D**) EDCs extracted at $\overline{\Gamma}$ for each doping condition. The color of the curves indicates corresponding ARPES data. The intensity is manually adjusted to give the best overlap.

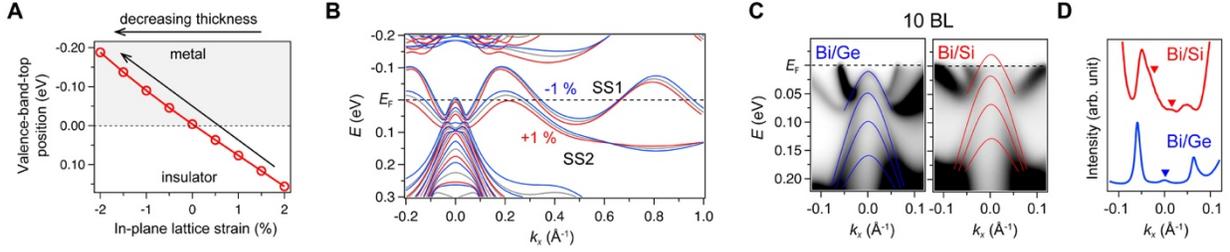

**Fig. S3. A finite effect caused by lattice strain from substrates.** (**A**) In-plane-strain dependence of the valence-band-top position obtained by first-principles calculations for the hole pocket of Bi, where the volume of a unit cell is kept constant. The arrow depicts a tendency of lattice distortion when a film thickness is reduced on a Ge substrate. (**B**) In-plane-strain dependence of band structures of a 14 BL Bi film obtained by tight-binding calculations. (**C**) Band structures measured on Bi films grown on Ge(111) and Si(111) substrates, respectively, with the same thickness of 10 BL. Solid curves are a guide to the eyes. (**D**) Momentum distribution curves extracted at $E_F$ in (**C**) with peak positions of the top quantized band highlighted.

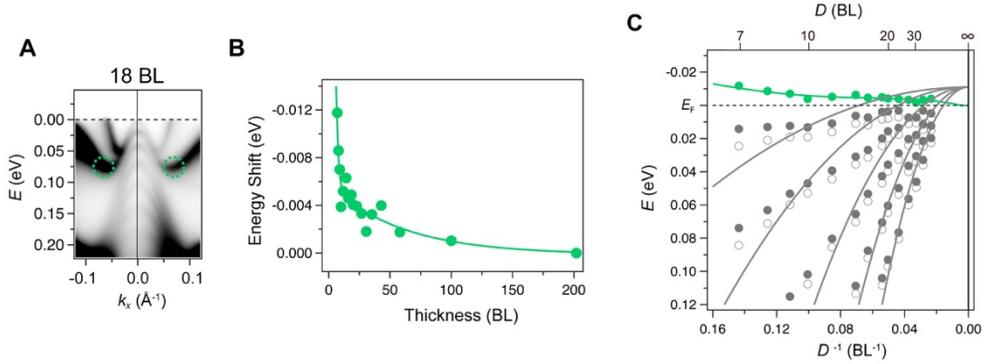

**Fig. S4. Experimental evaluation of minor contributions from all standard size effects.** (**A**) Band structures measured on a 18 BL Bi film (the same as in Fig. 1H of the main text). (**B**) Relative shifts of surface-state peak positions evaluated at the points highlighted by dashed circles in (**A**), where peak positions determined at positive and negative wavenumbers are averaged. A solid curve in (**B**) shows a double exponential fitting result. (**C**) The energy shifts in (**B**) plotted against inverse thicknesses (green marker and curve) along with the quantized energy levels in Fig. 2D of the main text (grey filled marker). Grey open markers show the energy levels corrected by the energy shifts. Grey solid lines depict examples of fitting results using parabolic functions.

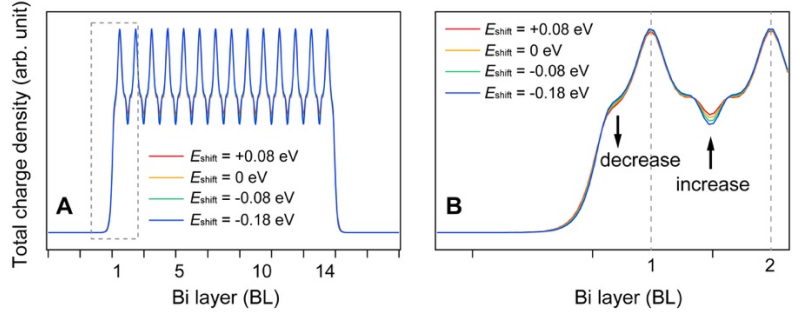

**Fig. S5. Changes of a total charge distribution in the first-principles calculations.** (**A**) Total charge densities calculated for the four cases in Fig. 3 (C and D) of the main text. The horizontal axis is shown in the same way as in Fig. 3D of the main text. (**B**) Total charge densities magnified inside a dashed box in (**A**). Decreasing and increasing tendencies respectively around slab edges and interiors are shown. Dashed lines denote bilayer positions.

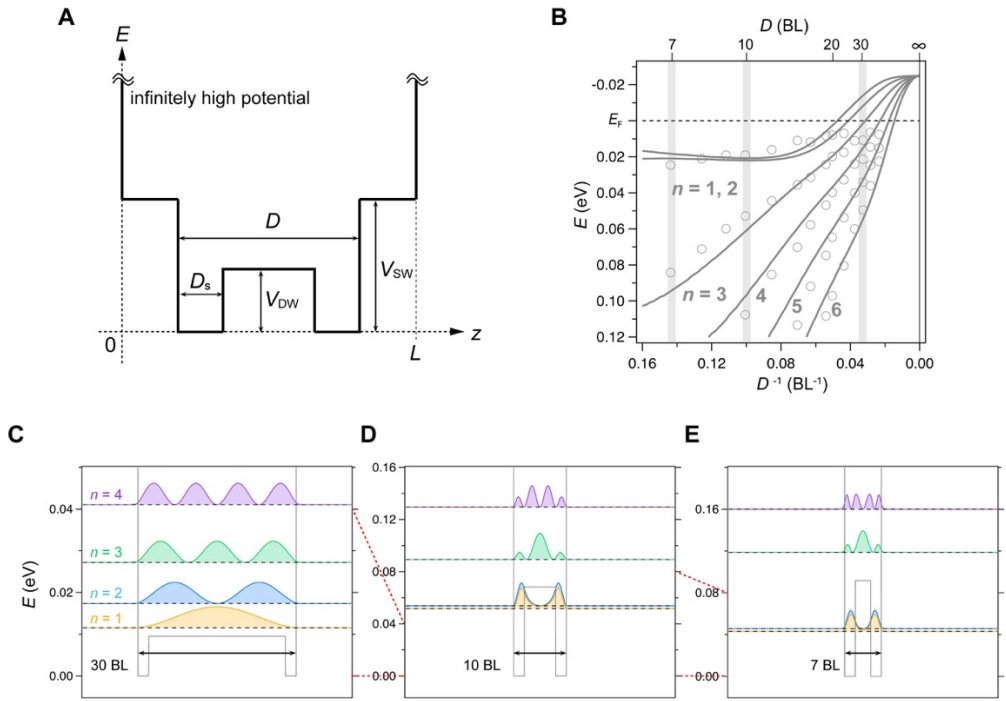

**Fig. S6. Details of model calculations in Fig. 4B of the main text.** (**A**) Schematic illustration of a 1D potential used in the numerical simulation. (**B**) Evolution of quantized energy levels simulated by the model calculation (solid curves) to best fit the experimental data corrected in Fig. S4C (open markers). Calculated energy levels are offset by -0.025 eV so that the values at the bulk limit coincide with the VBM of Bi at T point, which was determined by a semimetallic band overlap in ref. (*46*) and the CBM determined in ref. (*14*) (see also Section C). (**C**, **D**, and **E**) Squared values of wave functions plotted inside the potential at each quantized energy level for a 30 BL, 10 BL, and 7 BL cases. Note that energy scales become gradually smaller from (**C**) to (**E**), as highlighted by red dashed lines.